\documentclass[proof]{WileyASNA-v1}

\articletype{Article Type}%

\received{1 October 2020}
\revised{12 October 2020}
\accepted{25 October 2020}

\begin{document}

%!!!!!!!!!!!!!!!!!!!!!!!!!!!!!!!!!!!!!!!!!!!!!!
%!!!!!!!!!!! Title
%!!!!!!!!!!!!!!!!!!!!!!!!!!!!!!!!!!!!!!!!!!!!!!

\title{The quest for the quark-gluon plasma from the perspective of dynamical models of relativistic heavy-ion collisions}

%!!!!!!!!!!!!!!!!!!!!!!!!!!!!!!!!!!!!!!!!!!!!!!
%!!!!!!!!!!! Authors
%!!!!!!!!!!!!!!!!!!!!!!!!!!!!!!!!!!!!!!!!!!!!!!
\author[1,2,3]{Marcus Bleicher}
\author[4,2,1]{Elena Bratkovskaya}
\authormark{Marcus Bleicher and Elena Bratkovskaya}
%
%!!!!!!!!!!!!!!!!!!!!!!!!!!!!!!!!!!!!!!!!!!!!!!
%!!!!!!!!!!! Affiliations 
%!!!!!!!!!!!!!!!!!!!!!!!!!!!!!!!!!!!!!!!!!!!!!!
%
%
\address[1]{Institute for Theoretical Physics, Johann Wolfgang Goethe University, Max-von-Laue-Str. 1, 60438 Frankfurt am Main, Germany}
\address[2]{Helmholtz Research Academy Hessen for FAIR (HFHF), GSI Helmholtz	Center for heavy-ion Physics. Campus Frankfurt, 60438 Frankfurt, Germany}
\address[3]{ John von Neumann-Institut f\"ur Computing, Forschungzentrum J\"ulich,
52425 J\"ulich, Germany}
\address[4]{GSI Helmholtzzentrum für Schwerionenforschung GmbH, Planckstr. 1, 64291 Darmstadt, Germany}
\corres{GSI Helmholtzzentrum fuer Schwerionenforschung GmbH, Planckstrasse 1, 64291 Darmstad, Germany  -\email{E.Bratkovskaya@gsi.de}}

%------------------------------------------
%Please use the 'starred' version of the \texttt{abstract} command for typesetting the text of the online abstracts

\abstract{
The physics of heavy-ion collisions  is one of the most exciting and challenging directions of science for the last four decades. On the theoretical side one deals with a non-abelian field theory, while on the experimental side today's largest accelerators are needed to enable these studies. 
The discovery of a new stage of matter - called the quark-gluon plasma (QGP) - and the study of its properties is one of the major achievements of modern physics. 
In this contribution we briefly review the history of theoretical descriptions of 
heavy-ion collisions based on dynamical models, focusing on the personal experiences in this inspiring field.  
}

\keywords{Heavy-ion collisions, transport theory}

\maketitle
%-------------------------------------------------------
 
%\section{Introduction}
\section{Introduction to the history of dynamical modeling of heavy-ion collisions}
%
%%%%%%% Marcus %%%%%%%%%%
During the mid seventies and mid eighties of the last century, it was not a priori clear if high energetic heavy-ion reactions would allow to create super dense nuclear matter or if the reaction would become essentially transparent \cite{McLerran:1982cx,McLerran:1982cv}. Based on the parton picture \cite{Feynman:1978dt,Andersson:1983ia} or Bjorken's estimates \cite{Bjorken:1982qr} one assumed a large degree of transparency, while other groups using a hydrodynamic picture suggested the creation of shock waves that would lead to the creation of super dense matter \cite{Scheid:1974zz,Baumgardt:1975qv}.  
Later on, elaborate relativistic transport simulations became available but the controversy could not be settled based on theory alone. While parton-based approaches still favoured the idea of nuclear transparency, (see e.g. VENUS \cite{Werner:1988fq}), models based on the relativistic Boltzmann equation suggested a substantial stopping of the impinging nuclei, i.e. baryon number transport towards midrapidity even at high collision energies \cite{vonKeitz:1991kg,Bleicher:1996vq}.

Nowadays, these questions have fortunately been resolved by experimental data. Luckily, the colliding nuclei are not transparent over a wide range of collision energies. In fact even at relativistic collision energies the impinging ions have been shown to loose a substantial amount of energy leading to the creation of not only a hot, but also a very baryon rich matter. Only this fact has allowed the field to thrive and to use heavy-ion collisions as a laboratory for the exploration of neutron star matter and the properties of binary neutron star mergers and the equation of state of compact stellar objects.

As a matter of fact, the dynamics of the early universe in terms of the 'Big Bang' may be experimentally studied by ultra-relativistic nucleus-nucleus collisions at the Large-Hadron-Collider
(LHC) in terms of 'tiny bangs' at vanishing baryo-chemical potential. While at the GSI-Schwerionensynchrotron (SIS) or at the beam-energy scan with Relativistic Heavy-Ion Collider (RHIC's BES) program matter at moderate temperatures and high baryon densities can be explored and linked to the properties of stellar objects.

During the late 1980ies to the mid 1990ies both, hydrodynamic and transport theoretical descriptions of heavy-ion reactions where contenders for the best description of the Au+Au collision data taken at BNL's Alternating
Gradient Synchrotron (AGS) ($\sqrt {s_\mathrm{NN}}\approx 3-5$ GeV) and for the upcoming data from CERN's Super-Proton Synchrotron (SPS) colliding light systems and later Pb nuclei at a center of mass energy of about $\sqrt {s_\mathrm{NN}}\approx 20$ GeV.

Unfortunately, at that time hydrodynamical or hydro-ispired models where not able to describe the upcoming data to sufficient detail \cite{Sollfrank:1996hd,Schlei:1997hn,Mayer:1997mi}
On the contrary, hadronic transport theory provided an unprecedented accuracy and detailed information of the collision systems evolution. 
The relativistic transport approaches could be grouped in two categories: i) Relativistic Quantum Molecular Dynamics approaches, like RQMD, propagating the n-particle distribution \cite{Aichelin:1986wa,Hartnack:1989sd} and ii) Relativistic 
Boltzmann(Valsov)-\"Uhling-Uhlenbeck (usually abbreviated as RBUU or RVUU) approaches \cite{PhysRev.43.552}, propagating one-body distribution functions in averaged mean-fields \cite{Bertsch:1984gb,Aichelin:1985zz,Bertsch:1988ik,Cassing:1990dr}. 

During the early 1990's essentially a few active groups where spearheading the transport model era: 
\begin{itemize}
    \item 
The RQMD group in Frankfurt, using the 8-dimensional phase space approach developed by  Heinz Sorge and the group of Horst St\"ocker, who had merged the non-relativistic IQMD \cite{Aichelin:1986wa,Hartnack:1989sd} with 
 the LUND string model and implemented a large body of baryonic and mesonic resonances.  
    \item 
The RBUU group at Giessen, lead by Wolfgang Cassing and Ulrich Mosel. Later they extended the semi-classical BUU \cite{Cassing:1990dr} to covariant formulations and incorporated the LUND string model for the description of multi-particle  interactions in the expanding hadronic phase, which became the basis for the Hadron-String-Dynamics (HSD) transport approach \cite{Ehehalt:1996uq,Cassing:1999es}. 

   \item 
The group by Che-Ming Ko developing the 
 extended relativistic transport (ART - A Relativistic Transport) model \cite{Li:1995pra}  used 
for describing interactions among hadrons in the final hadronic phase.

\end{itemize}
%%%
%  --  microscopic models -----

\section{The need for partons and the deconfinement transition}

We recall that in the beginning microscopic transport models have been developed to explore the dynamics of hadrons using a nuclear matter EoS (where nuclear matter EoS usually meant some type of Skyrme potential and electric potential) and cross
sections based on measured experimental data or effective hadronic
Lagrangians. Later on, the models started to include the excitation and decay of color strings (essentially either based on the LUND picture (momentum exchange) or the parton picture based the idea of color exchange). Such improvements became necessary with the increase in collision energy which becomes essential at AGS and SPS energies.

\begin{figure}
\centerline{
\includegraphics[width=0.9\linewidth]{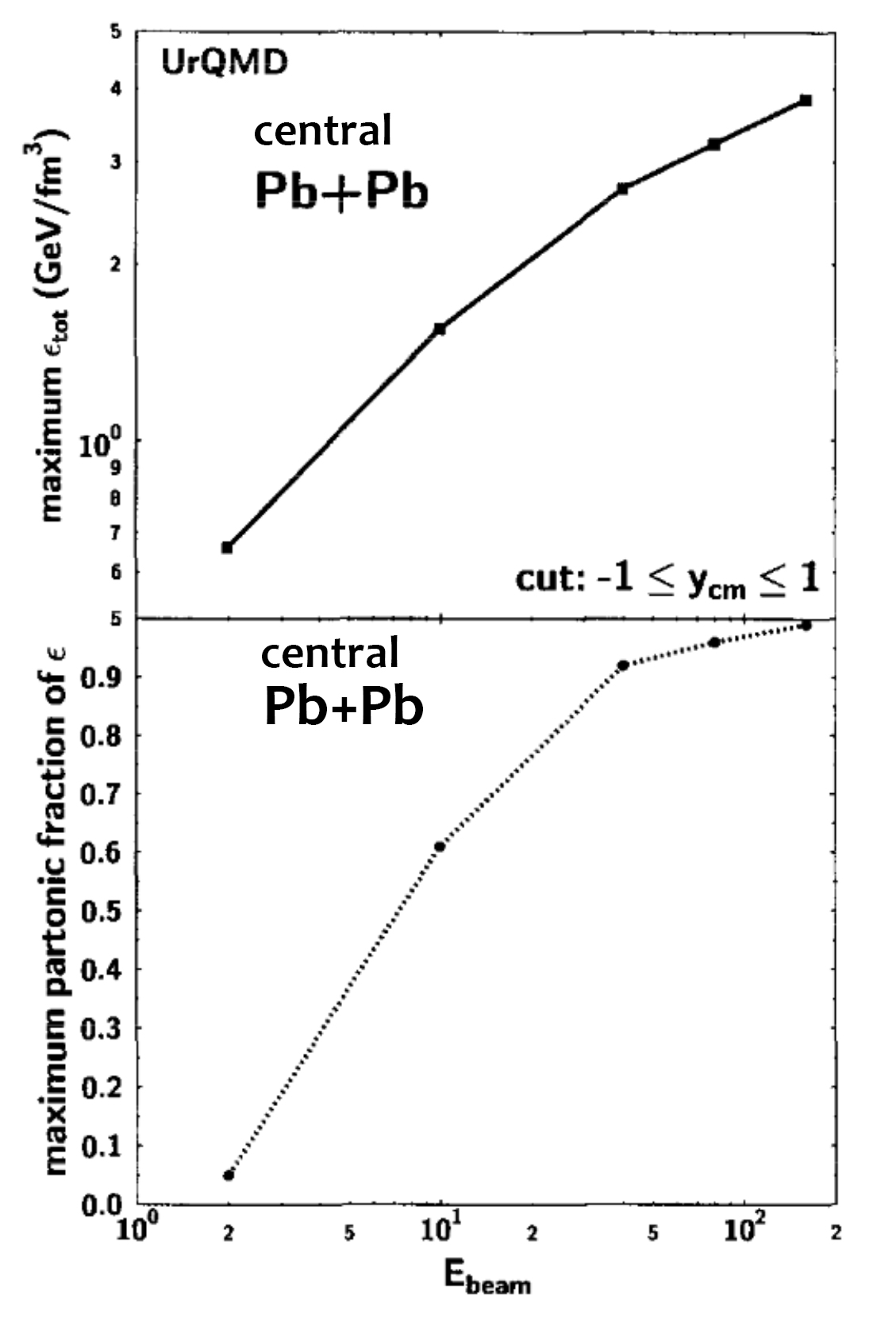}}
\caption{Top: Maximum energy density achieved in central Pb+Pb collisions at mid-rapidity. Bottom: Fraction of energy density in the 'quark state' (inside strings)  as a function of collision energy from UrQMD.
The figures are adopted from \cite{Weber:1998aa}.}
\label{Fig:partonfraction}
\end{figure}

%%%mb
Already at that time, first studies on the production of partons had been performed within these hadron-string models. E.g. in \cite{Weber:1998aa} it was shown that quark degrees of freedom - hidden inside strings - carry a substantial part of the energy density at collision energies above 30 AGeV. In Fig. \ref{Fig:partonfraction} we show the fraction of energy density in the quark state as a function of collision energy  in central Pb+Pb collisions at mid-rapidity as calculated within the UrQMD model. 
Here quark degrees of freedom mean those quarks that are produced during the string fragmentation process. Let us point out as a side remark: A precursor to the inclusion of full parton dynamics in the early stage of the reaction was the development of string fusion models, also called color ropes. Here, one recombines color charges from different string ends to higher color charges, which in turn modify the fragmentation of the color field (prime examples are \cite{Merino:1991nq,Amelin:1993cs,Armesto:1994yg,Sorge:1995dp}).

%%%

The observation of high parton densities in the UrQMD model (and also similar observations made in other models as discussed above) did already hint towards the need to implement partonic dynamics into transport simulations. However, it seemed that the influence of the partonic stage on observables in the SPS energy regime was rather moderate. Therefore, the immediate need to tackle this problem was not as pressing as it might look from todays perspective. This situation changed drastically when  the first RHIC data emerged. 

Among the first persons to implement an equation of state mimicking a phase transition between partonic and hadronic degrees of freedom could be implemented into transport simulation was Heinz Sorge \cite{Sorge:1998mk}. In contrast to todays implementation of the parton-hadron transition into simulations using mostly the real part of the potential to modify the EoS \cite{OmanaKuttan:2022the,Steinheimer:2022gqb}, Sorge used the virial theorem and implemented the QCD-EoS in the collision term. To show the effects emerging from the EoS with a phase transition in comparison to the hadron gas EoS, the elliptic flow of hadrons was suggested. As a matter of fact, elliptic flow has now become a central observable to map out the properties of the produced QCD matter.
%-------------------------

However, let us return to the timeline.  
When hadronic models where applied to heavy-ion collisions at RHIC energies a couple of problems emerged since a number of observables (elliptic flow of charged hadrons,
transverse mass spectra of hadrons, intermediate mass dileptons
etc.) could no longer be properly described by  hadron-string
degrees of freedom alone \cite{Bratkovskaya:2003ie,Bratkovskaya:2004kv}. 

%---------------------------------------------------------------------
%\section{Traces of QGP in observables -- problems of Hadronic transport models} %\label{s2}

\begin{figure}[h!]
\centering
\includegraphics[scale=0.17]{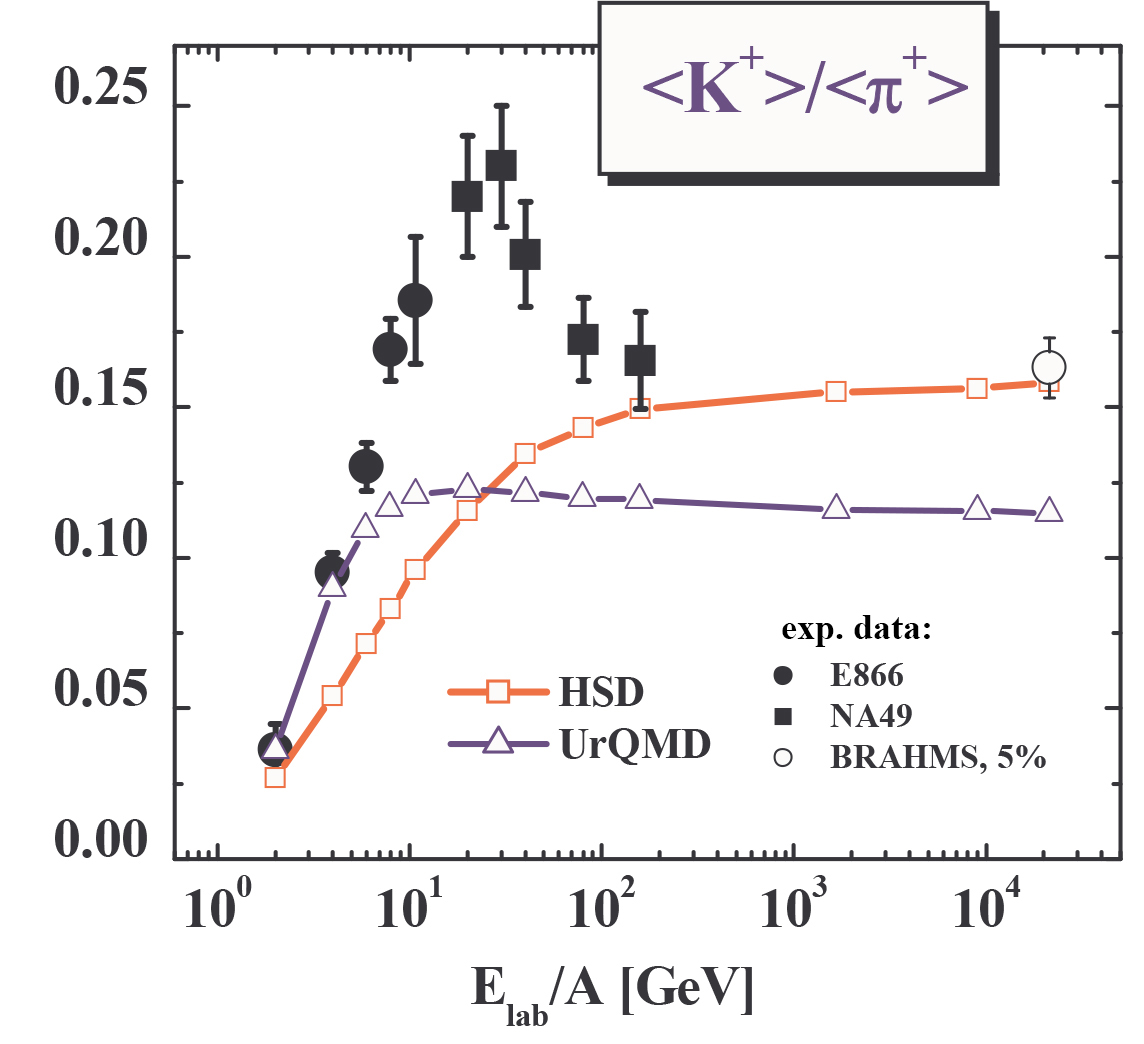}\\
\includegraphics[scale=0.165]{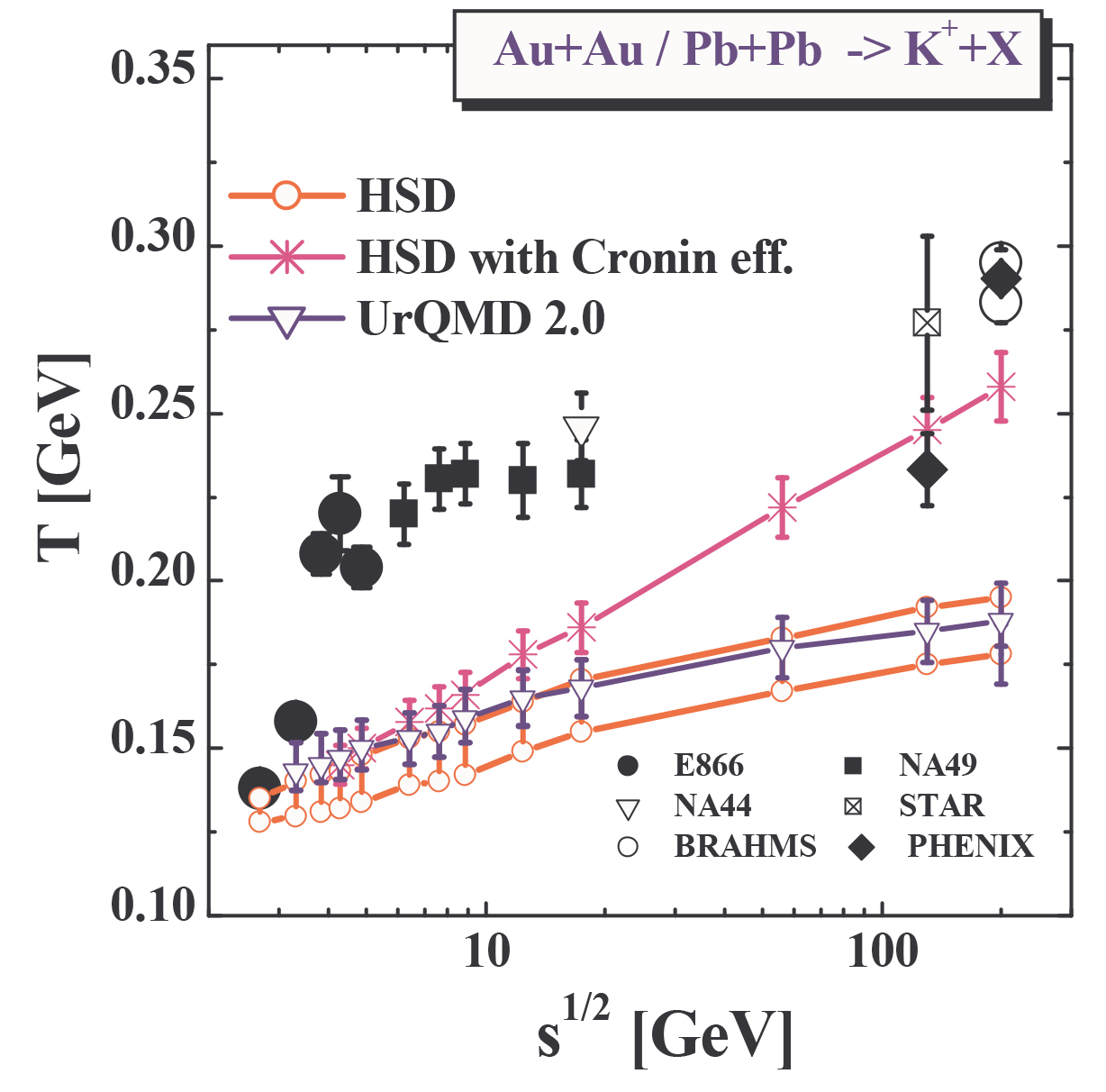} 
%\vspace*{5mm}
\caption{Excitation function of the $K^+/\pi^+$ ratio (l.h.s.) and inverse slope parameter for $K^+$ (r.h.s.) from central Au+Au (AGS and RHIC) or Pb+Pb (SPS) collisions.  The solid lines with open squares show the results from HSD whereas the dashed lines with open triangles indicate the UrQMD calculations. The solid lines with stars correspond to HSD calculations including 'Cronin' initial state enhancement. The figures are adopted from \cite{Bratkovskaya:2004kv}.} 
\label{Fig_KpiT} 
\end{figure}

One of the  observables which played an essential role in heavy-ion physics was the ratio of strange to non-strange hadrons.
As has been proposed in 1982 by Rafelski and M\"uller \cite{Rafelski:1982pu} the strangeness degree of freedom might play an
important role in distinguishing hadronic and partonic dynamics. In 1999  Ga\'zdzicki and Gorenstein \cite{Gazdzicki:1998vd} - within the
statistical model - have predicted  experimental observables which should show an anomalous behaviour at the phase transition: the
'kink' -- an enhancement of pion production in central Au+Au (Pb+Pb) collisions relative to scaled $pp$ collisions; the 'horn'
-- a sharp maximum in the $K^+/\pi^+$ ratio at 20 to 30 A$\cdot$GeV; the 'step' -- an approximately constant slope of
$K^\pm$ spectra  starting from 20 to 30 A$\cdot$GeV. Indeed, such "anomalies" have been observed experimentally by the NA49
Collaboration \cite{NA49:2002pzu,NA49:2003ksd}.

On the theoretical side we have investigated the hadron production as well as transverse hadron spectra in
nucleus-nucleus collisions from 2 $A\cdot$GeV to 21.3 $A\cdot$TeV within the independent transport approaches UrQMD and HSD \cite{Weber:2002pk,Bratkovskaya:2003ie,Bratkovskaya:2004kv}.  The comparison to experimental data demonstrates that both approaches
agree quite well with each other and with the experimental data on hadron production. The enhancement of pion production in central
Au+Au (Pb+Pb) collisions relative to scaled $pp$ collisions (the 'kink') is well described by both approaches without involving any
phase transition. However, the maximum in the $K^+/\pi^+$ ratio at 20 to 30 A$\cdot$GeV (the 'horn') is missed by $\sim$ 40\%
\cite{Weber:2002pk,Bratkovskaya:2004kv} -- cf. Fig. \ref{Fig_KpiT} (l.h.s.).  A comparison to the transverse mass spectra from $pp$ and C+C (or
Si+Si) reactions shows the reliability of the transport models for light systems \cite{Bratkovskaya:2003ie}. For central Au+Au (Pb+Pb)
collisions at bombarding energies above $\sim$ 5 A$\cdot$GeV, however, the measured $K^{\pm}$ $m_{T}$-spectra have a larger
inverse slope parameter than expected from the calculations. The approximately constant slope of $K^\pm$ spectra at SPS (the
'step') is not reproduced either \cite{Bratkovskaya:2003ie,Bratkovskaya:2004kv} -- cf. Fig. \ref{Fig_KpiT} (r.h.s.). 
%The HSD calculations also demonstrate that the 'partonic' Cronin effect plays a minor role at AGS and SPS energies for the  parameter $T$.  
The slope parameters from $pp$ collisions (r.h.s. in Fig.  \ref{Fig_KpiT}) are seen to increase smoothly with energy both in the experiment (full squares) and in the transport calculations (full lines with open circles) and are significantly lower than those from central Au+Au reactions for $\sqrt{s} > 3.5$ GeV.
Thus, the pressure generated by hadronic interactions in the transport models above $\sim$ 5 A$\cdot$GeV is lower than observed in the
experimental data. This finding suggests that the additional pressure - as expected from lattice QCD calculations at finite quark chemical
potential and temperature - might be generated by strong interactions in the early pre-hadronic/partonic phase of central Au+Au (Pb+Pb) collisions.

Another prime example showing the effect of the early partonic pressure was advanced by the competition between the groups performing hydrodynamic calculations \cite{Kolb:2000fha,Huovinen:2001cy} and transport simulations \cite{Bleicher:2000sx,Molnar:2001ux} on the description of elliptic flow. The main result was that the description of the elliptic flow data needs an extremely low viscosity 
%(or large transport opacity) 
of the matter created at RHIC. This was first recognized in \cite{Bleicher:2000sx} and later confirmed in more detail in \cite{Molnar:2001ux}. Of course the ideal hydrodynamic simulation done  at that time used $\eta/s=0$ by definition, allowing them to obtain an excellent description of the flow data at RHIC.

%----------------------------------------------------------------

Thus, the discrepancies between the experimental data and the results of existing (at those time) hadron-string based microscopic transport models stimulated an outstanding development of dynamical models.
The need to reformulate the transport models for partonic degrees of freedom became imminent. This included purely perturbative QCD (pQCD) based models, like Boltzmann-type simulations employing perturbative QCD (pQCD) derived two- and three-particle cross sections: Especially the pioneering work by Geiger (realized in the VNI model \cite{Geiger:1991nm}) popularized this approach, later followed by Zhangs Parton Cascade \cite{Zhang:1998hi} and Molnar's Parton Cascade \cite{Molnar:2000jh}. The pinnacle of parton cascades was reached with the realization of BAMPS \cite{Xu:2004mz} that included three gluon interactions in a consistent fashion and was extensively used to study the approach to thermalization of super hot gluon matter. However, all of these approaches did only employ an ideal gas EoS for the partonic phase and could not describe any phase transition.
Later on, some of these models where actually merged together to increase the region of applicability. Here the main representative of the modularized approaches is known as the ''A Multi Phase Transport Model'' (AMPT) \cite{Lin:2004en}, which merged Zhang's Parton Cascade and the ART transport model.

Unfortunately, pQCD scattering cross sections between massless partons turned out too low in order to describe the elliptic flow of hadrons
measured experimentally \cite{Molnar:2001ux,Xu:2008av}, either effective (enhanced) two-body cross sections have been used \cite{Ko:2012lhi} or additional $2 \leftrightarrow 3$ channels needed to be added as in BAMPS \cite{Xu:2004mz}. The formation of hadrons (if implemented at all) was usually performed by coalescence either in momentum space or - more recently - in phase space \cite{Ellis:1995pe,Ellis:1995fi}. Another branch of transport models is based on NJL-like approaches including a coupling to a scalar mean field and/or a
vector mean field  \cite{Florkowski:1995ei,Ruggieri:2013bda,Marty:2012vs}. In these models the partons have a finite dynamical mass and the binary cross sections are either extracted from the NJL Lagrangian \cite{Marty:2012vs} or parameterized to simulate a finite $\eta/s$ (as in hydro models)
\cite{Ruggieri:2013ova}. All these approaches provide a reasonable description of experimental data at RHIC energies as well as for LHC energies.

\subsection{Inclusion of a partonic phase  in transport models: The PHSD way}

In order to achieve a fully microscopic description of the hadronic and partonic phase,
the Parton-Hadron-String Dynamics (PHSD) transport approach has been developed \cite{Cassing:2008sv,Cassing:2008nn}. The PHSD is a microscopic covariant off-shell dynamical model for strongly interacting systems formulated on the basis of Kadanoff-Baym equations \cite{Cassing:2008nn} for Green's functions in phase-space representation (in first order gradient expansion beyond the quasiparticle approximation). The approach consistently
provides the full evolution of a relativistic heavy-ion collision from the initial hard scatterings and string formation through the
dynamical deconfinement phase transition to the strongly-interacting quark-gluon plasma (sQGP) as well as hadronization and the
subsequent interactions in the expanding hadronic phase as in the Hadron-String Dynamics (HSD) transport approach \cite{Ehehalt:1996uq,Cassing:1999es}.
\begin{figure*}[h!]
	\centering
 \phantom{a}\hspace*{10mm}\includegraphics[width=0.6\linewidth]{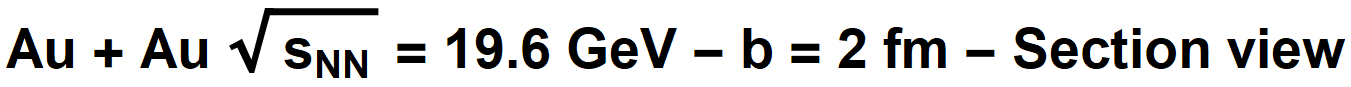} ~\\ \vspace{0.2cm}
 \phantom{a}\hspace*{5mm}\includegraphics[width=0.92\linewidth]{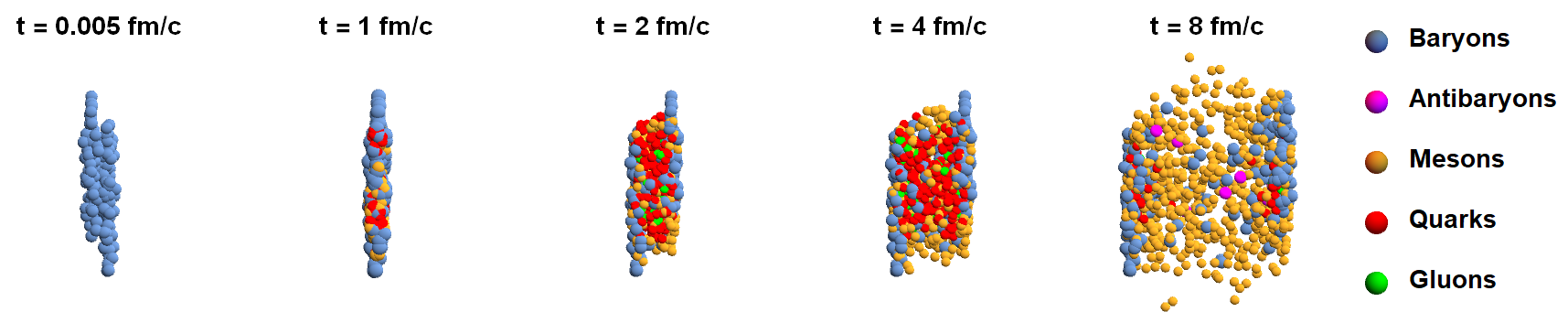} 	~\\ \vspace{0.2cm}
\includegraphics[width=0.85\linewidth]{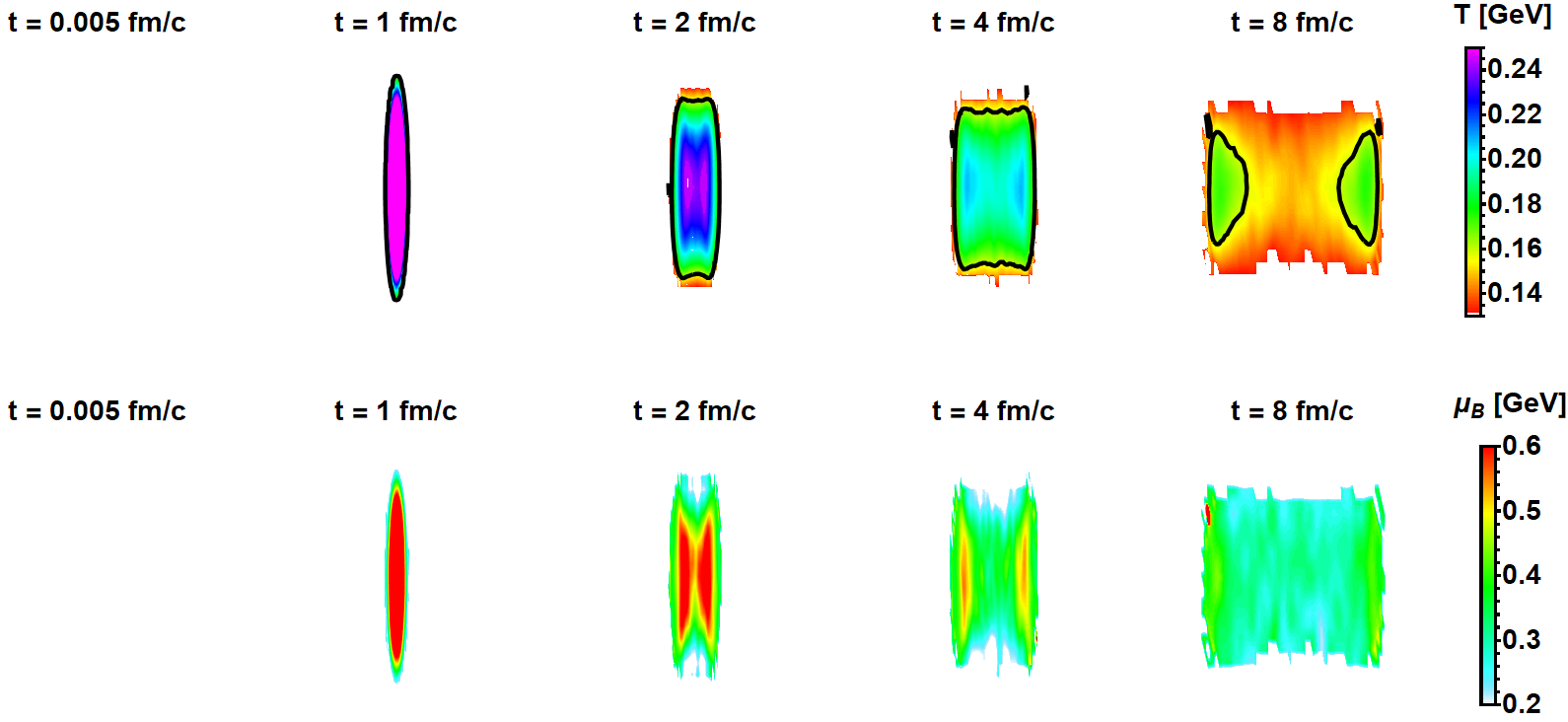}		
\caption{Illustration of the time evolution of central Au+Au collisions (upper row, section view) at a collisional energy of $\sqrt{s_{NN}} = 19.6$ GeV within the PHSD 
\cite{Moreau:2019aux}. 
The local temperature $T$ (middle row), baryon chemical potential $\mu_B$ (lower row),  as extracted from the PHSD for $y \approx 0$. 
The black lines (middle row) indicate the critical temperature $T_c \simeq 0.158$ GeV.
The figures are adopted from \cite{Moreau:2021clr}.}
	\label{Fig_HIC}
\end{figure*}

The transport theoretical description of the QGP phase in the PHSD is based on the Dynamical Quasi-Particle Model (DQPM) \cite{Peshier:2005pp,Cassing:2007nb,Cassing:2007yg,Berrehrah:2014kba} which is an effective model constructed to describe the strongly interacting non-perturbative nature of the QCD matter and to reproduce lattice-QCD results \cite{Aoki:2009sc,Cheng:2007jq,Borsanyi:2015waa} for a quark-gluon plasma in thermodynamic equilibrium.
The degrees-of-freedom in the DQPM are strongly interacting massive 
quarks and gluons  (contrary to the massless pQCD partons). They are described in terms of effective propagators with complex self-energies: the real part of self-energies corresponds to the effective finite masses (scalar mean-fields) for gluons/quarks while
the imaginary part - for the  finite widths related to the medium dependent reaction rate.
For fixed thermodynamic temperature $T$ and baryon chemical potential
$\mu_B$ the partonic width's $\Gamma_i(T,\mu_B)$ fix the effective two-body interactions represented in terms of quasi-elastic and inelastic parton cross sections that are implemented in the PHSD \cite{Ozvenchuk:2012fn}.

The parton scattering cross sections are probed by transport coefficients (correlators) in thermodynamic equilibrium by performing the PHSD calculations in a finite box with periodic boundary conditions (shear- and bulk viscosity, electric conductivity,  magnetic susceptibility etc. \cite{Ozvenchuk:2012kh,Cassing:2013iz,Steinert:2013fza}). 
The transport coefficients are of great importance for a 'viscous' hydrodynamical description of heavy-ion collisions since they are input parameters which define the 'deviation' of the QGP or hadronic 'fluid' from an ideal fluid \cite{Huovinen:2001cy,Song:2007fn,Romatschke:2007mq,Luzum:2008cw}.  

We note that in the recent versions of the PHSD 5.0 (and extensions)
the off-shell partonic interaction cross sections have been evaluated based on the leading order scattering diagrams and depend on $T, \mu_B$, the invariant energy of the colliding partons $\sqrt{s}$ as well as the scattering angle \cite{Moreau:2019vhw}.
In Refs. \cite{Moreau:2019vhw,Soloveva:2019xph,Fotakis:2021diq,Soloveva:2021quj} the transport coefficients (such as specific shear and bulk viscosities to entropy ratios $\eta/s, \zeta/s$, electric conductivities $\sigma_{QQ}/T$, diffusion coefficients etc.) 
have been explore in the $T,\mu_B$ plane based on relaxation-time approach and Kubo formalism and a good agreement with available lattice-QCD data has been found.

The PHSD successfully describes the many observables from SIS to LHC energies
for the $p+A$ and $A+A$ collisions
\cite{Cassing:2008sv,Cassing:2008nn,Cassing:2009vt,Bratkovskaya:2011wp,Linnyk:2015rco,Moreau:2019vhw}. For the theoretical foundation of the PHSD model we refer the reader to Ref. \cite{Cassing:2021book}.

In order to illustrate how the microscopic description of heavy-ion collisions proceeds within the PHSD, we show  in Fig. \ref{Fig_HIC} the time evolution  of central Au+Au collisions (upper row, section view) at a collisional energy of $\sqrt{s_{NN}} = 19.6$ GeV within the PHSD \cite{Moreau:2019aux}. 
The snapshots are taken at times $t=0.005, 1, 2, 4$ and 8 fm/c.
The baryons, antibaryons, mesons, quarks and gluons  are shown as colored dots.
The middle row of Fig. \ref{Fig_HIC} shows  the local temperature $T$ and the lower row displays the baryon chemical potential  $\mu_B$, as extracted from the PHSD in the region with $y \approx 0$. 
The black lines (middle row) indicate the critical temperature $T_c = 0.158$ GeV.
As follows from the upper part of Fig. \ref{Fig_HIC}, the QGP is created 
in the early phase of the collisions and when the system expands, a hadronization occurs.
One can see that during the overlap phase the temperature $T$ and chemical potential $\mu_B$ are very large
and then decrease with time. However, even at 8 fm/c there are "hot spots" of QGP
at the front surfaces of high rapidity.

%-------------- hybrid UrQMD 
\subsection{Inclusion of partonic phase  in transport models: The UrQMD hybrid way}

An alternative way to include a phase transition into a transport simulation is to couple a relativistic Boltzmann-equation to the hydrodynamic equations. By this, one is able to describe the initial non-equilibrium dynamics of the QCD matter and its thermalization, while this state then provides a source term for the hydrodynamic evolution in which a QCD equation of state with a (phase) transition between a hadron gas and a QGP can be included. After the phase transition, when local equilibrium can not be maintained any longer, one uses a loss term in the hydrodynamic equations and transfers the matter back into a relativistic Boltzmann simulation. This allows to model the freeze-out/decoupling stage using resonant hadronic scattering from the loss of local equilibrium (chemical freeze-out) to kinetic freeze-out \cite{Steinheimer:2017vju}. In this way one can merge the advantages of hydrodynamical and Boltzmann-type simulations.
Physically, such an approach is possible due to the sufficiently strong parton interactions, leading to rather fast equilibration in the central collision
zone and allowing to keep  local equilibrium after some
initial delay. It is further supported by the success of plain (ideal) hydrodynamical models \cite{Ollitrault:1992bk,Heinz:2001xi,Shuryak:2008eq} in description 
of the first  experimental data on the elliptic flow $v_2(p_T)$ at RHIC.

This lead to the development of hybrid models
which incorporate three different type of model components:
i) the initial nonequilibrium phase to specify the initial state fluctuations or initial flow;
ii) viscous or ideal hydrodynamic for the hadronic and partonic (fluid) phase including a possible phase transition;
iii) hadronic 'afterburner' for resonant interactions in the hadronic phase to model the freeze-out after the loss of local equilibrium. 

Due to the matching of the different phases a couple of new parameters enter such models that define the matching conditions.
Accordingly, a multi-parameter approach (on the scale of $\sim$ 15 independent parameters) emerges that has to be optimized in 
comparison to a multitude of experimental data in order to extract physical information on the transport coefficients. This has been
done within a Bayesian analysis by a couple of authors and some proper information could be extracted so far on $\eta/s(T)$ as well
as for the charm diffusion coefficient $D_s(T)$ \cite{Bass:2017zyn,Auvinen:2017fjw,Xu:2017obm}. For explicit results we refer the reader
to Refs. \cite{Bass:2017zyn,Auvinen:2017fjw,Xu:2017obm,Bernhard:2016tnd,Pratt:2016lol,Sangaline:2015isa,Pratt:2015zsa}.
We note in passing that within such approaches semi-central and central nucleus-nucleus collisions at ultra-relativistic energies can well
be described \cite{Song:2013qma} but an application to elementary high-energy $p+p$ or $\pi+p$ reactions is restricted to very high energies and high multiplicity triggers \cite{Werner:2013tya}.

A pioneering framework in this class of models is the ultra-relativistic quantum molecular dynamics (UrQMD) hybrid approach which starts with
UrQMD  \cite{Bass:1998ca,Bleicher:1999xi} for the initial nonequilibrium phase on an event by event basis, switches to hydro after the baryon currents have separated from each other in phase space and approximate equilibration in the local cells is reached.  The evolution continues with a hydrodynamic expansion until a critical energy density is reached in each cell (individually) and merges, after a Cooper-Fry particlization,  into the UrQMD cascade to describe the final hadronic rescatterings \cite{Steinheimer:2007iy,Petersen:2008dd}. By construction such hybrid models may be used for lower (AGS) energies as well as for ultra-relativistic (LHC) energies. Further improvements by incorporating e.g. a color glass condensate (CGC), IP-glasma or EPOS2 initial conditions \cite{McDonald:2018wql,Gelis:2016upa,McDonald:2020oyf,Schenke:2014jza,Werner:2010aa,Werner:2012xh,Nahrgang:2014ila}  provide also a very good
description of the collective flows as a function of bombarding energy and collision centrality at RHIC and LHC energies.

%----------------------------------- UrQMD vs PHSD 
\subsection{QGP in theoretical laboratories - UrQMD(-hybrid) and PHSD}

Looking in the past and having the present experience we believe that it was a fortunate decision by our transport groups to follow  alternative ways in incorporating a quark-gluon plasma in our transport approaches UrQMD and PHSD: \\
{\it i)} to follow the fully microscopic description of QGP by the inclusion of partonic degrees of freedom and their interaction in the PHSD explicitly and \\
{\it ii)} to follow a hybrid description by switching from the microscopic hadronic phase to a macroscopic description of the QGP phase in terms of a propagating 'fluid'.

Each of these approaches has advantages and disadvantages. The basic differences are the following:  the advantage of the PHSD is a fully microscopic description of all stages of heavy-ion collisions - partonic and hadronic - with full conservation of energy-momentum and all quantum numbers  in each elementary collision, which allows to analyze the  history of each individual parton or hadron created during the expansion of the system.  
In the UrQMD it is not possible to do so due to the switch to a 'fluid' description for the partonic phase by converting the energy density and flow velocity of individual cells to the hydro evolution, which leads to a discontinuity in entropy density etc. However, in a hydrodynamical description it is much more straightforward to incorporate and study different equations of state, e.g. a crossover lattice-EoS at zero $\mu_B$ or some chiral model EoS with a 1st order phase transition. Moreover, one avoids an explicit solution of the hadronization problem and replaces it by the 'particalarization' problem which is much easier in realization using e.g. the Cooper-Frye procedure. In the PHSD the incorporation of different EoS requires first it's interpretation in terms of degrees of freedom and their interactions. Since lattice-QCD can not provide this information and delivers only averaged thermodynamic quantities, one needs to develop some effective models - as  done by the PHSD group when the DQPM model has been introduced. Moreover, since the QGP is a strongly interacting matter, one can not apply semi-classical BUU-type of kinetic approaches which are valid for the description of systems where the mean-free pass is larger then the range of interactions. The description of strongly interacting quasiparticles with  dynamical broad spectral functions and complex self-energies required a substantial step in the development of the kinetic
transport approach based on a field-theoretical description within Kadanoff-Baym theory and a derivation of new transport equations - i.e. the Cassing-Juchem equations for the test particles in first-order gradient expansion \cite{Juchem:2004cs,Cassing:2008nn} - to be solved numerically. 

Finally,  two comprehensive transport approaches - the UrQMD(-hybrid) and PHSD - 
have been developed and they became real theoretical 'laboratories' to study the dynamics of heavy-ion collisons on  computers.

%---------------------------------------------------------------------
\section{Success of transport models with the QGP} %\label{s2}
%------------------- fraction of QGP
\begin{figure}[t!]
	\centering
	\includegraphics[width=8cm]{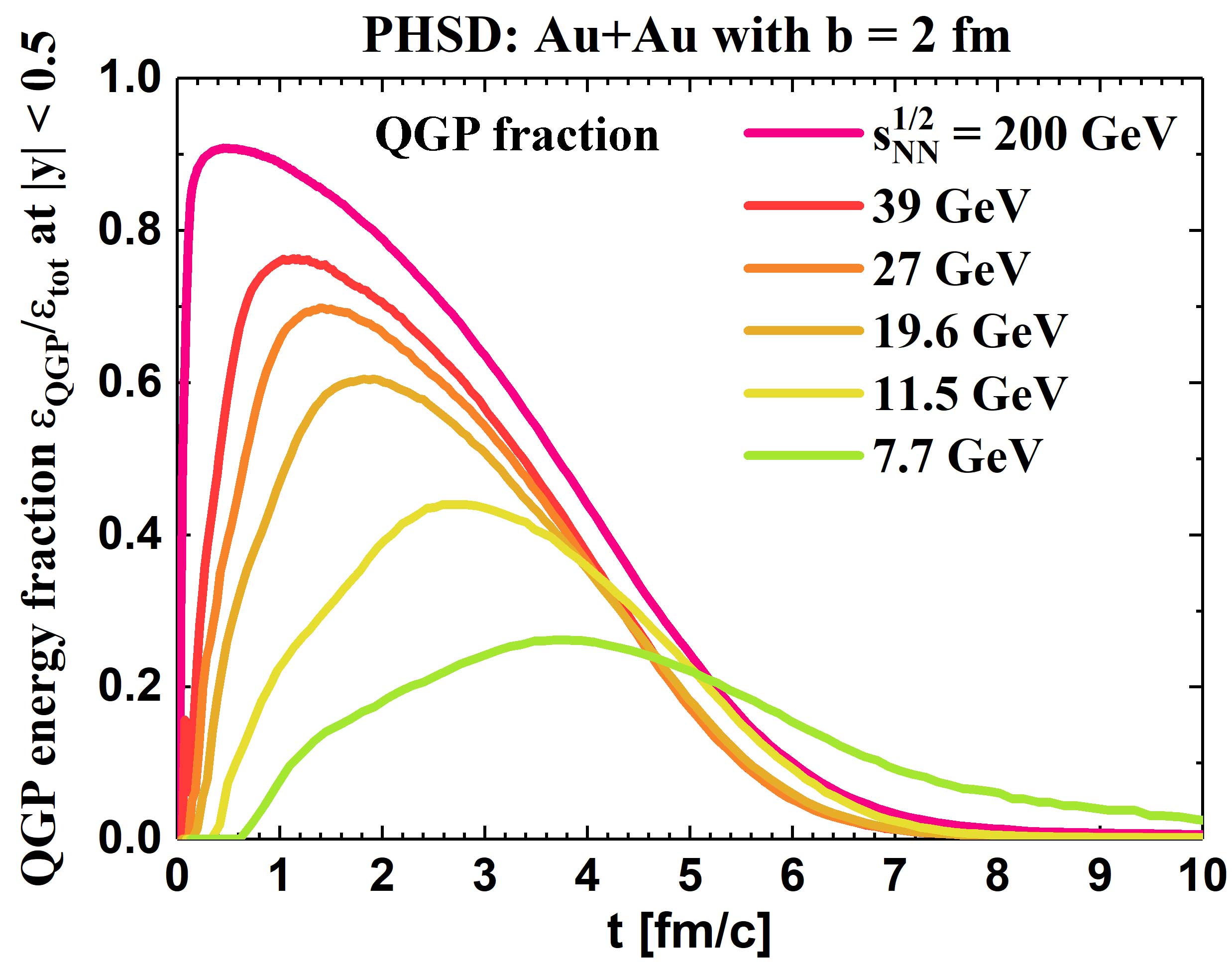} 
	\caption{The QGP energy fraction from PHSD  as a function of time $t$ in central (impact parameter b = 2 fm) Au+Au collisions for different collisional energies $\sqrt{s_{NN}}$, taking into account only the midrapidity region $|y| < 0.5$.  
The figures are adopted from \cite{Moreau:2021clr}.} 
	\label{Fig_QGPfrac}
\end{figure}

\begin{figure*}[!h]
\centerline{
\includegraphics[width=0.85\linewidth]{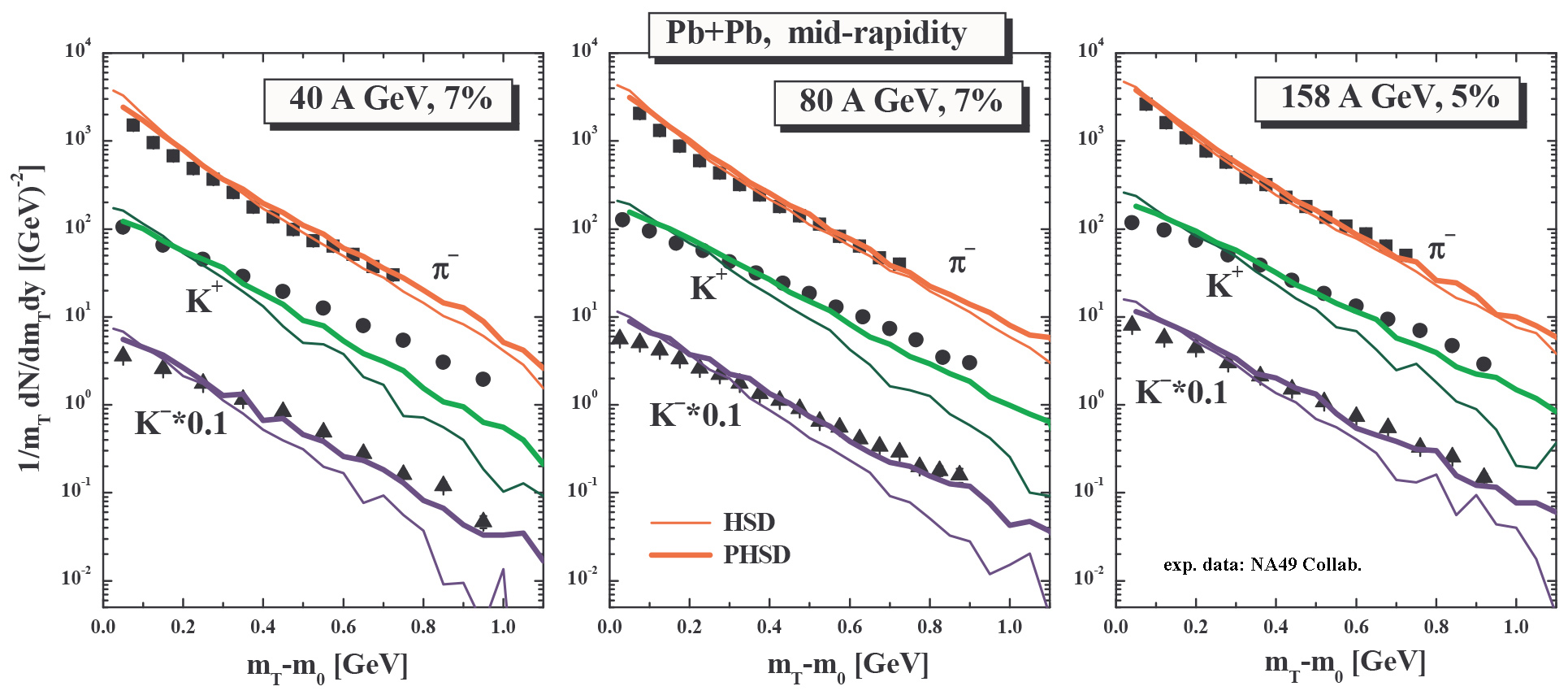}}
\caption{The $\pi^-$, $K^+$ and $K^-$ transverse mass spectra for
central Pb+Pb collisions at 40, 80 and 158 A$\cdot$GeV from PHSD (thick
solid lines) in comparison to the distributions from HSD (thin solid
lines) and the experimental data from the NA49 Collaboration
\cite{NA49:2002pzu,NA49:2007stj}. The figures are adopted from 
\cite{Cassing:2009vt}.}
\label{Fig_mtkpi}
\end{figure*}

  %------------------------ v2
\begin{figure}[h!]
\centerline{
\includegraphics[scale=0.09]{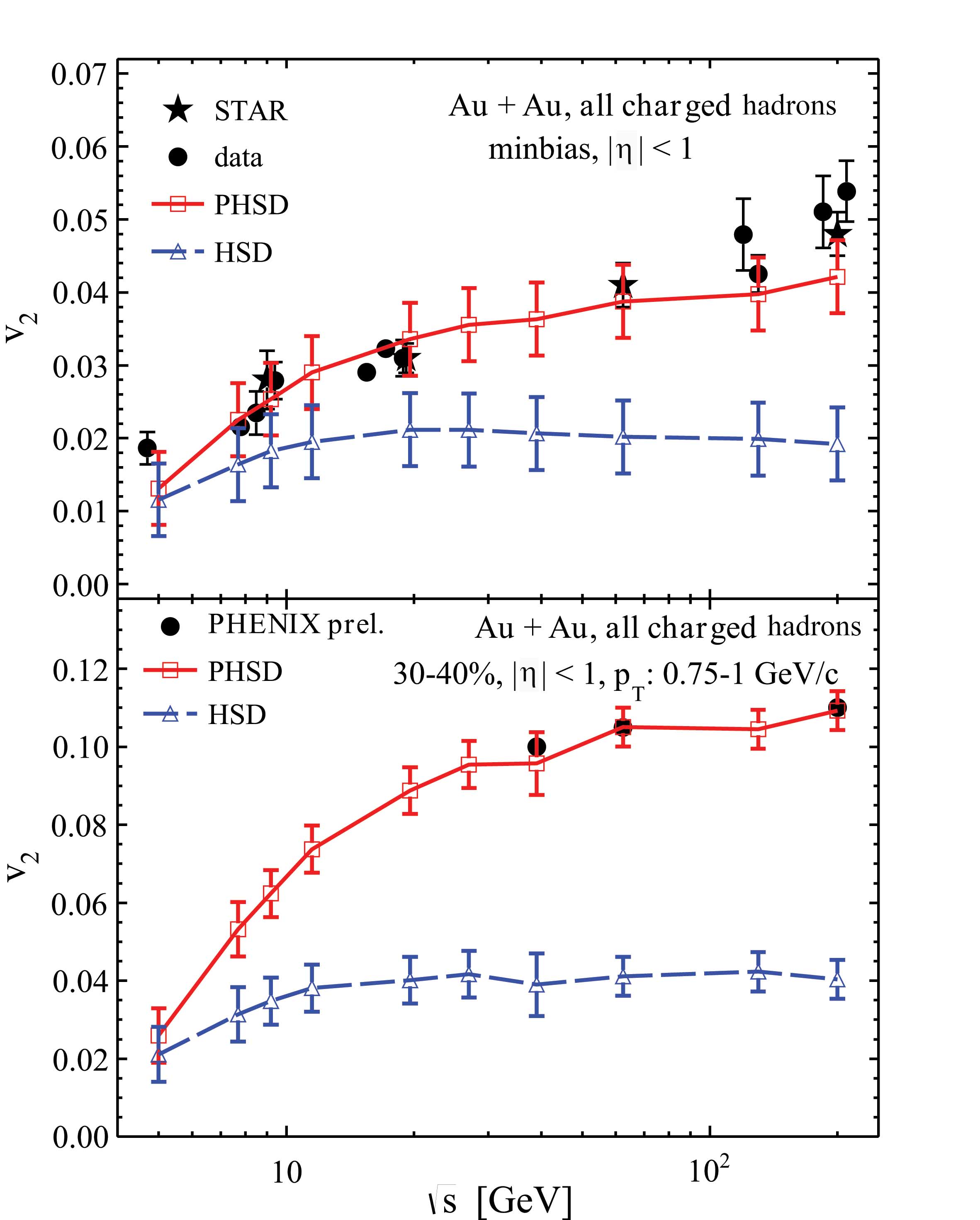}}
\caption{ Average elliptic flow $v_2$ of charged particles at   mid-pseudorapidity for two centrality selections calculated within   the PHSD (solid curves) and HSD (dashed lines) models. 
  The experimental data for minimal bias are  from the STAR Collaboration (stars) \cite{Nasim:2010hw},  from  the PHENIX Collaboration (full circles) \cite{Gong:2011zz} and other data are taken from the compilation in   Ref.~\cite{STAR:2009sxc}.
  The figures are adopted from \cite{Konchakovski:2012yg}.  }
\label{vns}
\end{figure}

\begin{figure}[h!]
\centerline{
\includegraphics[width=0.95\linewidth]{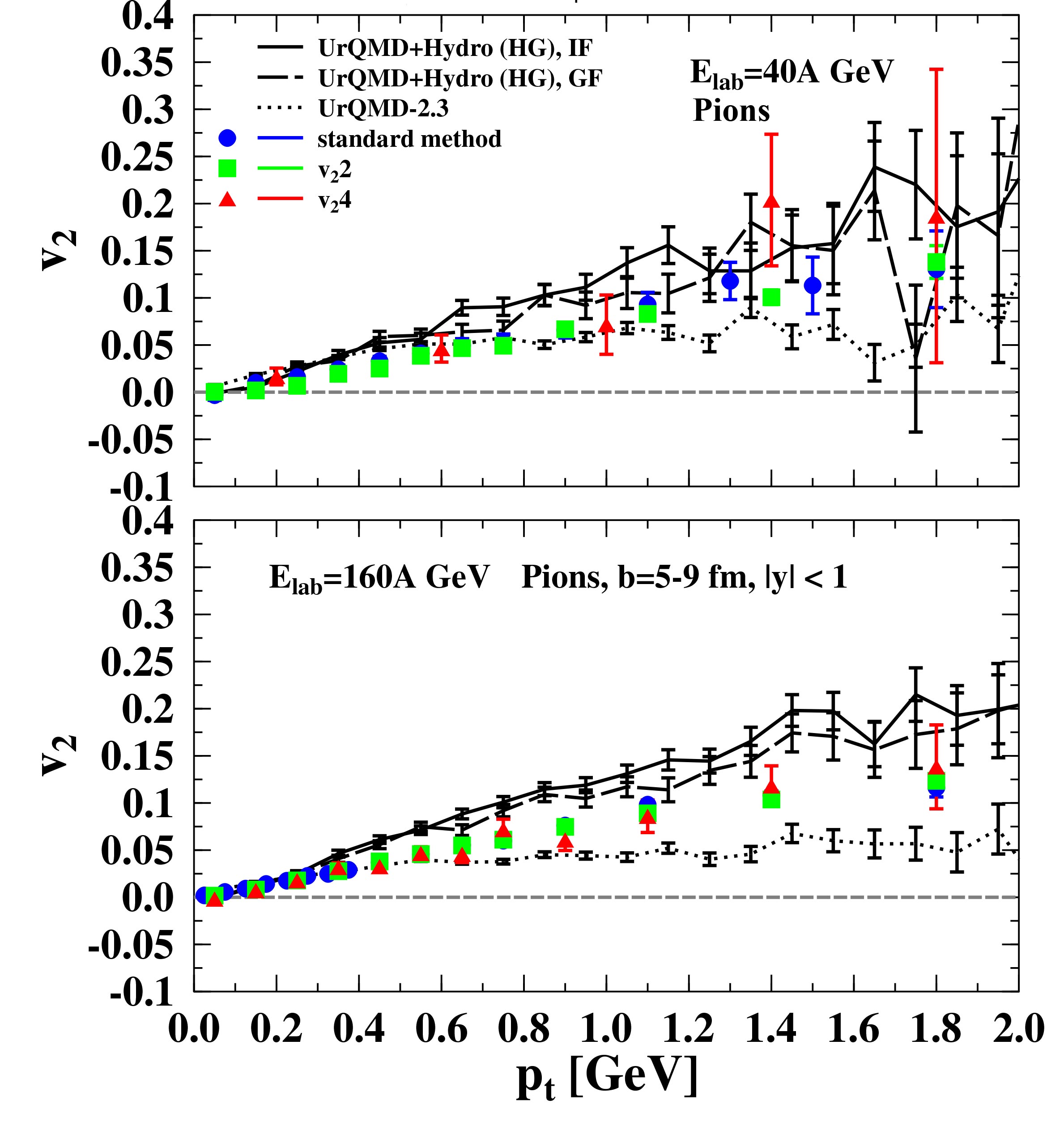}}
\caption{Transverse momentum dependence of the elliptic flow $v_2$ of pions from Au+Au collisions at mid-rapidity for two beam energies ($40A$GeV and $160A$GeV) from UrQMD. Full and dashed lines show the results of the hybrid simulation, while the dotted lines show the result of the standard hadronic cascade mode. The data by NA49 is shown by symbols.
  The figures are adopted from \cite{Petersen:2015rra}.  }
\label{elliptic_flow_comparison}
\end{figure}

In this section we recall a few most prominent examples how the  explicit inclusion of the QGP dynamics - on a fully microscopic level in transport approaches or via the hydrodynamic phase in the hybrid approach - substantially improves the  description of experimental observables at relativistic energies. For a more detailed comparison we refer the reader to the review \cite{Bleicher:2022kcu}.

It is important to stress that the influence of the QGP increases with increasing bombarding energy.
Although QGP droplets can be created already at relatively low collision energies of 3 - 5 GeV  due to local fluctuations in energy density, their size is tiny and their traces in observables are hardly visible since the dynamics is dominated by hadronic interactions.
Oppositely, with increasing collision energy of heavy-ions the QGP fraction grows such that partonic interactions become dominant for about 10 fm/c in Au+Au collisions and have  visible consequences on observables. 

In Fig. \ref{Fig_QGPfrac} we show the QGP energy fraction versus the total energy for Au+Au at different collisional energies $\sqrt{s_{NN}}$ accounting only for the midrapidity region $|y| < 0.5$. One can see that for high energies the QGP fraction is  large compared to lower collisional energies where the QGP volume is small.  
While in high energy heavy-ion collisions the QGP phase appears suddenly after  the initial primary NN collisions, at low energies its appearance is smoother  since the passing time of the nuclei (inversely proportional to the Lorentz $\gamma$-factor of A+A collision) is much longer. Correspondingly, at low energies the QGP lifetime is large, however, the total QGP volume is very small; thus its influence on the dynamics is much reduced compared to high energy collisions where practically 90\% of matter at midrapidity is in the QGP phase  (at least for a short time).  We note that the PHSD calculations for the partonic fraction in Fig. \ref{Fig_QGPfrac} are qualitatively consistent with the early UrQMD estimates shown in Fig. \ref{Fig:partonfraction}. 

We recall that, as discussed in Section 1.2, one of the main problems with the hadron-string dynamical description of heavy-ion collisions is related to a 'missing pressure' in transverse direction reflected in such observables as $p_T$- (or $m_T$-) spectra and  elliptic flow $v_2$. This has been illustrated in Fig. \ref{Fig_KpiT} (r.h.s.) which shows a substantial deviation of the inverse slope parameter $T$ of the transverse spectra in comparison to experimental data. Since in heavy-ion collisions a pressure generation can occur only by interactions, this clearly demonstrates that hadronic interactions are not sufficient to push the matter in transverse direction, i.e. partonic interactions are needed. Now we step on to the same observables but include a QGP phase in the microscopic transport approach PHSD.

In Fig. \ref{Fig_mtkpi} we show the $\pi^-$, $K^+$ and $K^-$ transverse mass spectra for central Pb+Pb collisions at 40, 80 and 158 A$\cdot$GeV from PHSD (thick solid lines) in comparison to the distributions from HSD (thin solid lines) and the experimental data from the NA49 Collaboration
\cite{NA49:2002pzu,NA49:2007stj}. 
One can see that the calculations in the PHSD mode (i.e. including the formation of the sQGP) provide a better description of the experimental data compared to the HSD one (i.e. in the string-hadron mode).

%-------------v2 ------------------
%-------------------------- K+/pi+
\begin{figure}[h!]
\begin{center}
\includegraphics[width=0.4\textwidth]{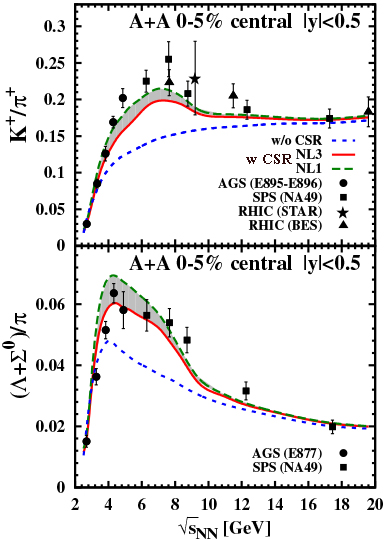} 
\end{center}
\vspace{-0.1cm}\caption{The ratios $K^+/\pi^+$ and
$(\Lambda+\Sigma^0)/\pi$ at midrapidity from 5\% central Au+Au
collisions as a function of the invariant energy $\sqrt{s_{NN}}$ up
to the top SPS energy in comparison to the experimental data.  
The grey shaded area represents the
results from PHSD including chiral symmetry restoration (CSR) taking into account the uncertainty
from the parameters of the $\sigma-\omega$-model for the nuclear
EoS. The figure is adopted from Ref. \cite{Palmese:2016rtq}.}
 \label{CSRhorn}
\end{figure}

\begin{figure}[h!]
\centerline{
\includegraphics[width=8.cm] {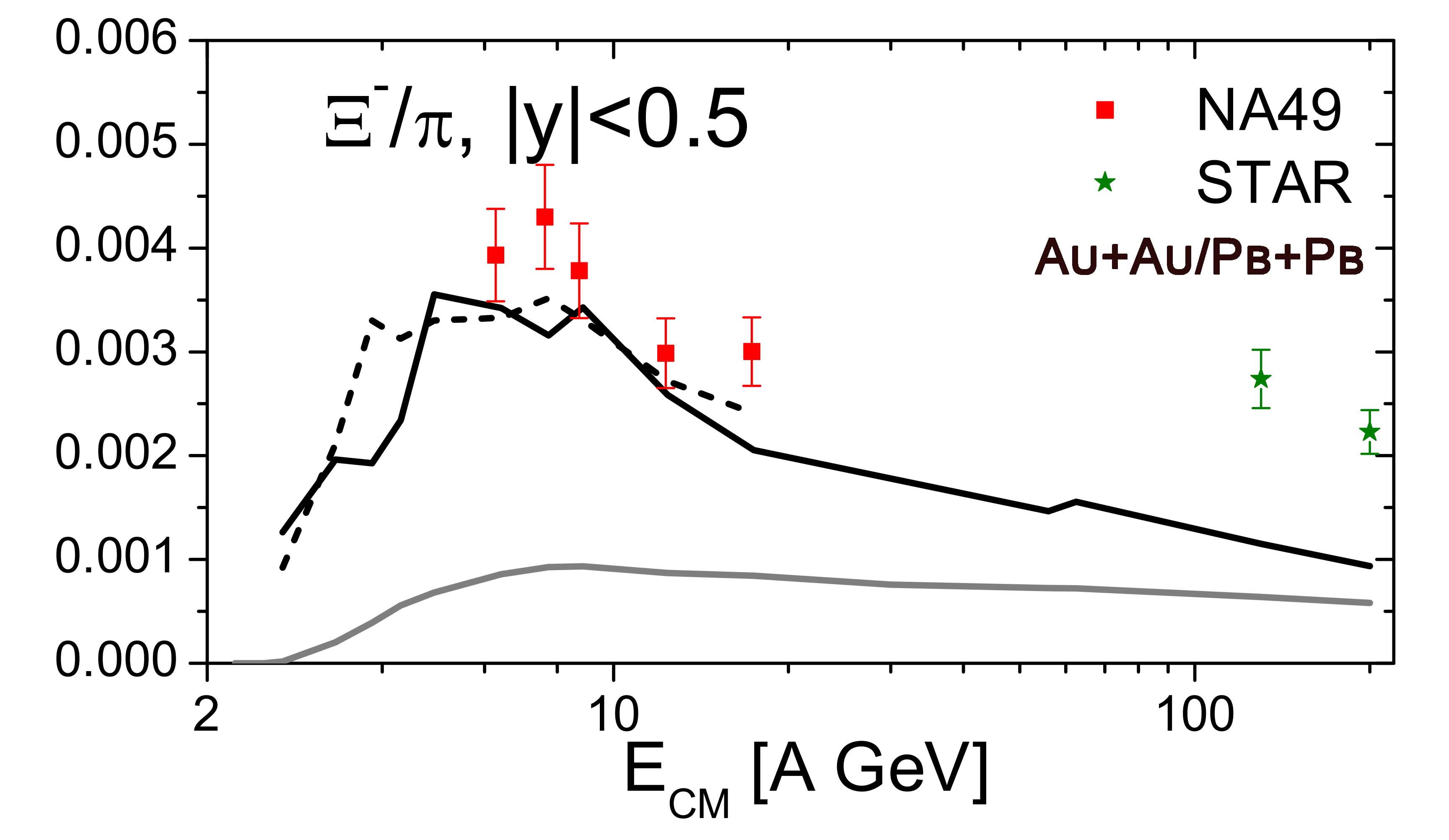}}
 \caption{ Energy dependence of $\Xi^-/\pi$ ratios  in Au+Au/Pb+Pb collisions. Black lines show results of the hybrid model with a phase transition, the gray line shows the UrQMD result without phase transition.
  The figure is adopted from Ref. \cite{Steinheimer:2010zza}}
\label{Fig_uhbrid_xi}
\end{figure}

Now we continue with  the next example for partonic interactions in the QGP phase that play an important role in the building of pressure in the system. This can be probed by the elliptic flow coefficient $v_2$ which is a widely
used quantity characterizing the azimuthal anisotropy of emitted
particles,
 \begin{eqnarray}
 \label{eqv2}
 v_2 = <cos(2\psi-2\Psi)>=<\frac{p^2_x - p^2_y}{p^2_x + p^2_y}>~,
 \end{eqnarray}
where $\Psi_{RP}$ is the azimuth of the
reaction plane, $p_x$ and $p_y$ are the $x$ and $y$ component of the particle
momenta and the brackets denote averaging over particles
and events. The azimuthal anisotropies have been studied within the PHSD in Ref. \cite{Konchakovski:2012yg} and here we show one of the prominent examples from this study.
In Fig, \ref{vns} we show the average elliptic flow $v_2$ of charged particles at   mid-pseudorapidity for two centrality selections calculated for  two cases: i) the PHSD model (solid curves), i.e. including the QGP formation and partonic interactions in terms of the DQPM, and ii) the HSD model (dashed lines) -- without formation of the QGP, i.e. within the hadron-string dynamics.  
One can see that the pure hadronic scenario is not able to describe the experimental data for the excitation function of  $v_2$. 
The errorbars indicate the statistical fluctuations in the PHSD results since the calculation of $v_2$ is a computer time consuming task, because the signal is of the order of few percent only.

Similar finding has been done with the UrQMD model \cite{Petersen:2015rra}:
in Fig. \ref{elliptic_flow_comparison} we show the transverse momentum dependence of the elliptic flow $v_2$ of pions from Au+Au collisions at mid-rapidity for two beam energies ($40A$GeV and $160A$GeV) from UrQMD. Full and dashed lines show the results of the hybrid simulations, which include the QGP phase in EoS, while the dotted lines show the result of the standard hadronic cascade mode (i.e. without the QGP interactions).
%(see also Fig. \ref{elliptic_flow_comparison} for the elliptic flow comparison using UrQMD). 

We note that the deviation of hadron-string models (HSD and UrQMD)  from the data grows with the energy, which clearly indicates that the partonic interactions (as implemented in a microscopic way in the PHSD or in the hybrid UrQMD)  are getting more and more important with increasing energy.
Indeed,  the volume of the QGP grows with increasing energy and the QGP phase becomes dominant at top RHIC energies as illustrated in Fig. \ref{Fig_QGPfrac}.

Figs. \ref{Fig_mtkpi} and \ref{vns} provide an example for the  importance of the partonic degrees of freedom in the dynamical description of the heavy-ion collisions. We note that the incorporation of the QGP in the PHSD left the $K^+/\pi^+$ ratio puzzle unsolved \cite{Cassing:2015owa} - the PHSD was still missing the 'horn' structure at $~\sim 20$ AGeV. Indeed, the QGP volume is rather low at this energies - cf. Fig. \ref{Fig_QGPfrac} and can not substantially change the 'chemistry' of the reactions, i.e. enhance the strangeness production. On the other hand, the hadron density in the early 'cold' stage of heavy-ion collisions - when the hadron production occurs via string formation and decay - is very high. 
Thus, one should expect such phenomena as a signal of partial chiral symmetry restoration (CSR) on the quark level inside strings. In Refs. \cite{Cassing:2015owa,Palmese:2016rtq} the description of chiral symmetry restoration via the Schwinger mechanism for the string decay  in a dense medium has been incorporated in the PHSD. This leads to a dropping of the 'dressed' quark masses - coupled to the scalar quark condensate  which changes the final chemistry of produced hadrons during the string breaking. In  the transport model the scalar quark condensate can be estimated via the scalar density of baryons and mesons based on the non-linear $\sigma$-$\omega$-model \cite{Friman:1997sv}.

Figure \ref{CSRhorn} shows the ratios $K^+/\pi^+$ and $(\Lambda+\Sigma^0)/\pi$ at midrapidity from 5\% central Au+Au collisions as a function of the invariant energy $\sqrt{s_{NN}}$ up to the top SPS energy in comparison to the experimental data. The grey shaded area represents the results from PHSD including chiral symmetry restoration  taking into account the uncertainty from the parameters of the $\sigma-\omega$-model for the nuclear EoS. As compared to the blue dashed lines, which show the 'old' HSD results without CSR and QGP (as shown in the left part of Fig. \ref{Fig_KpiT}), one sees a clear enhancement of the ratios due to the enhanced strangeness production in the initial phase of the collisions.  Thus, the inclusion of chiral symmetry restoration  together with a partonic phase allows to describe the maximum in the $K^+/\pi^+$ ratio as an interplay between the dense hadronic medium and the QGP transition. 

In Fig. \ref{Fig_uhbrid_xi} we show the energy dependence of $\Xi^-/\pi$ ratios  in Au+Au/Pb+Pb collisions. Black lines show results of the hybrid model with phase transition, the gray line shows the UrQMD result without phase transition.

For a variety of further observables, which signal the formation of the QGP, we refer the reader to the recent review \cite{Bleicher:2022kcu}.

%-------------------------------------------

\section{Final remarks}

In this review we have recalled the historical progress in the  dynamical modeling of heavy-ion collisions which rapidly developed on the boarder of the new millennium and was strongly driven by  experimental observations. Since the QGP can not be observed directly in experiments, a proper theoretical  interpretation of the experimental measurements is mandatory.  This can be done in a consistent way only by transport approaches - derived from  ab-initio kinetic and many-body theories and providing a full description of the time evolution of heavy-ion collisions, following all stages of the expanding system. An important step in our understanding of experimentally measured results has been achieved due to the theoretical development of  transport approaches by inclusion of the quark-gluon plasma phase: {\it i)} on a fully microscopic level by means of partonic degrees of freedom and their explicit interactions as in the PHSD or {\it ii)}  on a macroscopic level by means of a hydrodynamical description of the QGP fluid based on a partonic equation-of state for the QGP phase. 

We have demonstrated for some prominent examples that the partonic phase is mandatory for a proper
description of the experimental observables on particle yields, ratios, spectra
as well as on the flow harmonics characterizing the dynamical expansion during the  relativistic heavy-ion collisions. Furthermore, we have shown that the strangeness degrees of freedom show an exceptional 
sensitivity to the dynamical description of the heavy-ion collisions.

Our historical examples indicate the importance of comprehensive efforts of theory and experiments in obtaining  progress in our understanding of many physical phenomena happening in nature.

%!!!!!!!!!!!!!!!!!!!!!!!!!!!!!!!!!!!!!!!!!!!!!!
%!!!!!!!!!!! References
%!!!!!!!!!!!!!!!!!!!!!!!!!!!!!!!!!!!!!!!!!!!!!!
%\bibliographystyle{ieee}
\bibliography{biblio}%
\end{document}